\documentclass[12pt]{article}
\usepackage{amsmath,amssymb}
\usepackage{bm}
\usepackage{color}
\usepackage{graphicx}
\usepackage{cite}
\usepackage{mathtools}
\usepackage{breqn}
\usepackage{braket, multirow, subfigure, multicol}
\usepackage{here}
\usepackage{slashed}
\usepackage{ulem}

\newcommand{\maru}[1]{\raise0.2ex\hbox{\textcircled{\scriptsize{#1}}}}

\begin{document}
\title{
\begin{flushright}
\ \\*[-80pt]
\begin{minipage}{0.2\linewidth}
\normalsize
EPHOU-23-019\\*[50pt]
\end{minipage}
\end{flushright}
{\Large \bf
Radiative correction on moduli stabilization \\ 
in modular flavor symmetric models
\\*[20pt]}}

\author{
~Tatsuo Kobayashi$^{a}$, 
~Kaito Nasu$^{a}$, 
~Riku Sakuma$^{a}$, and
~Yusuke Yamada$^{b}$
\\*[20pt]
\centerline{
\begin{minipage}{\linewidth}
\begin{center}
{\it \normalsize
${}^{a}$Department of Physics, Hokkaido University, Sapporo 060-0810, Japan\\
${}^{b}$Waseda Institute for Advanced Study, Waseda University, 1-21-1 Nishi Waseda, Shinjuku, Tokyo 169-0051, Japan} \\*[5pt]
\end{center}
\end{minipage}}
\\*[50pt]}

\date{
\centerline{\small \bf Abstract}
\begin{minipage}{0.9\linewidth}
\medskip
\medskip
\small
We study the radiative corrections to the stabilization of the complex structure modulus $\tau$ in modular flavor symmetric models. 
We discuss the possibility of obtaining the vacuum expectation value of $\tau$ in the vicinity of the fixed point where residual symmetries remain unbroken. As concrete examples, we analyze the 1-loop Coleman-Weinberg potential in the $A_4$ modular flavor models. We show that the 1-loop correction may lead to the slight deviation from the tree level result, which may realize a phenomenologically preferred value of the complex structure modulus $\tau$ particularly when the number of species contributing to the 1-loop correction is large enough.
\end{minipage}
}
\begin{titlepage}
\maketitle
\thispagestyle{empty}
\end{titlepage}
\newpage


\section{Introduction}
\label{Intro}

One of important issues to study in particle physics is to understand the origin of flavor structure.
Quark and lepton masses are hierarchical.
The lepton mixing angles are large, while the quark mixing angles are small.
Modular flavor symmetry is one of attractive approaches to 
the flavor mysteries in particle physics \cite{Feruglio:2017spp}.
Indeed, modular flavor symmetric models have been studied 
extensively. (See for earlier works Refs.~~\cite{Kobayashi:2018vbk,Penedo:2018nmg,Criado:2018thu,Kobayashi:2018scp,Novichkov:2018ovf,Novichkov:2018nkm,deAnda:2018ecu,Okada:2018yrn,Kobayashi:2018wkl,Novichkov:2018yse} and for more references Ref.~\cite{Kobayashi:2023zzc}.)

The modular symmetry is a geometrical symmetry of compact space such as torus.
The modulus $\tau$ of the torus transforms non-trivially under the modular symmetry.
Yukawa couplings as well as other couplings and masses are 
functions of the modulus in four-dimensional (4D) low energy effective field theory derived from higher dimensional theory such as superstring theory.
Thus, Yukawa couplings as well as others transform non-trivially 
under the modular symmetry.
Indeed Yukawa couplings are modular forms.
Matter fields and Higgs fields also transform non-trivially.
They can be representations of finite modular groups such as 
$S_3$, $A_4$, $S_4$, $A_5$, etc.
The 4D effective field theory must be invariant under the 
modular symmetry including non-trivial transformations of 
Yukawa couplings and other couplings.
Thus, concrete values depend on the vacuum expectation value (VEV) of the modulus.
Indeed, the modulus stabilization was studied within 
the framework of modular flavor models Refs.~\cite{Kobayashi:2019xvz,Ishiguro:2020tmo,Abe:2020vmv,Novichkov:2022wvg,Ishiguro:2022pde,Knapp-Perez:2023nty,King:2023snq}.

Generically, the modular symmetry is broken completely when the modulus value is fixed.
Some residual symmetries remain at certain fixed points.
The $Z_2$ and $Z_3$ symmetries remain at $\tau = i$ and $\tau = \omega = e^{2\pi i/3}$, respectively, while $T$ symmetry remains in the limit $\tau = i \infty$.
The modular forms behave like $Y \sim \varepsilon^n$ around the fixed point, where $\varepsilon \ll 1$ and $n$ denotes the charge of matter under the residual symmetry.
Such behavior is very interesting to quark and lepton mass hierarchies 
without fine-tuning \cite{Feruglio:2021dte,Novichkov:2021evw,Petcov:2022fjf,Kikuchi:2023cap,Abe:2023ilq,Kikuchi:2023jap,Abe:2023qmr,Petcov:2023vws,Abe:2023dvr,deMedeirosVarzielas:2023crv}.
On the other hand, some mechanisms stabilize the modulus value 
at an exact fixed point, e.g. $\tau=\omega$ \cite{Ishiguro:2020tmo,Abe:2020vmv}.
Such stabilization at an exact value does not lead to realistic results 
in fermion mass matrices, but small deviation is useful from 
the phenomenological viewpoint.
However, we may have some corrections and the stabilized modulus value may shift slightly from the fixed point.
Such deviation would be useful to realize hierarchical masses as well as 
mixing angles, although the deviation depends on the ratio of corrections to the stabilized modulus mass.
Our purpose is to study radiative corrections on the modulus stabilization 
due to heavy modes through the Coleman-Weinberg potential.\footnote{Moduli stabilization only by radiative corrections was studied in Refs.~\cite{Kobayashi:2017aeu,Kobayashi:2017zfd}.}

This paper is organized as follows.
In section \ref{sec:Revew}, we give a brief review on modular flavor symmetric models and 
behavior of modular forms in the vicinity of fixed points.
In section \ref{sec:model}, we illustrate our $A_4$ models
for stabilizing the modulus and 
one-loop effective potential.
In section \ref{sec:results}, we study the potential numerically and analytically by some approximations.
Section \ref{conclusion} is our conclusion.
Group theoretical aspects of $A_4$ are reviewed in Appendix \ref{appendix: group_A4}.
$A_4$ modular forms are listed in Appendix \ref{sec: A_4}.
In Appendix \ref{appendix: approximation}, we show approximations, which are used in section 
\ref{sec:results}.


\section{Modular symmetry as a flavor symmetry}
\label{sec:Revew}
Here, we briefly review the 4D $\mathcal{N}=1$, modular-invariant supersymmetric(SUSY) model.
We consider the following infinite groups,
\begin{equation}
    \Gamma(N) = \left\{ 
    \begin{pmatrix}
    a & b \\ c  & d 
    \end{pmatrix}
     \in SL(2,\mathbb{Z}),\ 
    \begin{pmatrix}
    a & b \\ c  & d 
    \end{pmatrix}
      \equiv
    \begin{pmatrix}
    1 & 0 \\ 0  & 1 
    \end{pmatrix}
    \pmod{N} \right\} ,
\end{equation}
where $N$ is a positive integer. 
The action of $\gamma\in \Gamma(N)$
\begin{equation}
    \gamma \tau = \frac{a\tau + b}{c \tau +d},
\end{equation}
is called modular transformation,
under which the upper half-plane $\{\tau \in \mathbb{C}| {\rm Im}\tau > 0 \}$ is mapped to itself. Since the transformations generated by $\gamma$ and $-\gamma$ on $\tau$ are identified, one often defines $\bar{\Gamma}(N) \equiv \Gamma(N)/\{ \mathbb{I},-\mathbb{I} \}$. Note that  $\bar{\Gamma}(N>2) = \Gamma(N)$ because {$-{\mathbb{I}} \notin \Gamma(N>2)$.} 
The group $\bar{\Gamma}(1)$ consists of the following two generators 
\begin{equation}
    S = \begin{pmatrix}
    0 & 1 \\ -1 & 0
    \end{pmatrix},\quad 
    T= \begin{pmatrix}
    1 & 1 \\ 0 & 1
    \end{pmatrix},
\end{equation}
which induce 
\begin{equation}
   S: \tau \rightarrow -\frac{1}{\tau},\quad T: \tau \rightarrow \tau + 1. 
\end{equation} 
The finite modular subgroups $\Gamma_N$ are defined as $\Gamma_N \equiv \bar{\Gamma}(1)/\bar{\Gamma}(N)$. 
For $N \leq 5$, we have the following presentations,
\begin{equation}
\label{eq: representation}
    \Gamma_N = \langle S,T | S^2 = {\mathbb I} , (ST)^3 = {\mathbb I}, T^N = {\mathbb I} \rangle.
\end{equation}
It is known that $\Gamma_N$ is isomorphic to the non-Abelian discrete symmetry groups $S_3, A_4, S_4, A_5$ for $N=2,3,4,5$ respectively \cite{deAdelhartToorop:2011re}, 
 which we will identify as flavor symmetries. \footnote{
 See Refs.~\cite{Altarelli:2010gt,Ishimori:2010au,Kobayashi:2022moq,Hernandez:2012ra,King:2013eh,King:2014nza,Tanimoto:2015nfa,King:2017guk,Petcov:2017ggy} for model building with these non-Abelian discrete flavor symmetries.}

The matter chiral superfields $\Phi_I$ transform as ``weighted" multiplets 
\begin{equation}
 (\Phi_I)_i \rightarrow (c\tau + d)^{-k_I} \rho_I(\gamma)_{ij} (\Phi_I)_j, \quad \gamma \in \Gamma_N,
\end{equation}
where $k_I \in \mathbb{Z}$ and $\rho_I(\gamma)$ is a unitary representation of $\Gamma_N$.
The 4D ${\cal N}=1$ global SUSY Lagrangian invariant under the $\Gamma_N$ modular symmetry is given by
\begin{equation}
    \mathcal{L} = \int d^2 \theta d^2 \bar{\theta}\  K(\tau, \bar{\tau}, \Phi_I, \bar{\Phi}_I) + \left[ \int d^2\theta\  W(\tau,\Phi_I) + {\rm h.c.} \right],
\end{equation}
where $K$ and $W$ are the K\"{a}hler potential and superpotential, respectively. 
We consider the following K\"{a}hler potential
\begin{equation}
\label{eq: Kahler}
    K(\tau, \bar{\tau}, \Phi_I,\bar{\Phi}_I) = - \Lambda_0^2 \log{(-i(\tau - \bar{\tau}))} + \sum_I \frac{|\Phi_I|^2}{(- i (\tau - \bar{\tau}))^{k_I}},
\end{equation}
where $\Lambda_0$ is a mass parameter. The K\"ahler potential $K$ transforms under $\Gamma_N$, which can be identified as a K\"ahler transformation $K\to K+\log|\Lambda|^2$ where $\Lambda$ is a chiral superfield. Thus the superpotential $W$ must transform in such a way that the K\"ahler invariant function $G=K+\log|W|^2$ is invariant under the modular transformation. 
Such superpotential $W$ 
is constructed by extracting trivial singlet $\bf{1}$ terms under $\Gamma_N$ from the products of holomorphic functions $Y_{I_1...I_m}(\tau)$ and matter chiral superfields,
\begin{equation}
\label{eq: superpotential}
    W(\tau, \Phi_I) = \sum_m \sum_{
    \{ I_1,...,I_m \} }  \left(
    Y_{I_1...I_m}(\tau) \Phi_{I_1} ... \Phi_{I_m}
    \right)_{\bf{1}}.
\end{equation}
From the $\Gamma_N$ invariance of $W$ in Eq.\,(\ref{eq: superpotential}), we find that $Y_{I_1,...,I_m}(\tau)$ must be a modular form of level $N$ with a specific modular weight $k_Y$. 
The modular forms of weight $k_Y$ and level $N$ are defined as holomorphic functions of $\tau$ which behave under $\Gamma(N)$ as,
\begin{equation}
\label{eq: Gamma(N)_transformation}
    Y_{I_1...I_m}(\gamma \tau) = (c \tau +d)^{k_Y} Y_{I_1...I_m}(\tau),\quad   \begin{pmatrix}
    a & b \\ c  & d 
    \end{pmatrix} \in \Gamma(N),
\end{equation}
where $k_Y$ is a non-negative even number. 
Acting $\gamma \in \Gamma_N$ on the modular forms, we obtain
\begin{equation}
\label{eq: Gamma_N_transformation}
    [Y_{I_1...I_m} (\gamma \tau)]_i = (c \tau + d)^{k_Y} \rho(\gamma)_{ij} [Y_{I_1...I_m} (\tau)]_j,\quad \begin{pmatrix}
    a & b \\ c  & d 
    \end{pmatrix} \in \Gamma_N,
\end{equation}
where $\rho(\gamma)$ denotes a unitary representation matrix of $\gamma \in \Gamma_N$. {The derivation of Eq.\,(\ref{eq: Gamma_N_transformation}) from Eq.\,(\ref{eq: Gamma(N)_transformation}) is presented in Appendix B of Ref.~\cite{Feruglio:2017spp}. Here, we briefly comment on the relation between the two equations. By definition, any elements of $\Gamma(N)$ are identified as the identity element in the quotient group $\Gamma_N$. Thus, the corresponding representation matrix is the identity matrix $\rho = \mathbb{I}$, if $\Gamma(N)$ actions are considered in Eq.\,(\ref{eq: Gamma_N_transformation}). This recovers Eq.\,(\ref{eq: Gamma(N)_transformation}) as required.}
For the $\Gamma_N$ invariance of $W$, the modular weight must satisfy $k_Y = k_{I_1}+...+k_{I_m}$. In addition, the tensor product of the representations of superfields and modular forms need to contain a trivial singlet $\bf{1}$ to get a non-vanishing superpotential.

\subsection{Behavior of modular forms in the vicinity of $\tau = \omega$}
When the modulus $\tau$ acquires a VEV, the modular $\Gamma_N$ symmetry is broken in general. However, at the fixed points $\tau = i \infty, i, \omega = e^{\frac{2\pi i}{3}}$, residual symmetries $Z_N, Z_2$ and $Z_3$ remain respectively.
In fact, it is phenomenologically attractive to have VEVs close to the fixed points when reproducing hierarchical flavor structures without fine-tuning. This is due to the fact that the values of modular forms become hierarchical depending on the charges 
of the residual symmetry \cite{Novichkov:2021evw}.
Thus, we will 
focus on stabilizing the modulus in the proximity of those fixed points. As we see in the next section, the 1-loop effective potential of $\tau$  will be written in terms of modular forms. This motivates us to 
look at the behavior of modular forms in the vicinity of the fixed point. Here and hereafter, we focus on $\tau = \omega = e^{\frac{2\pi i}{3}}$ where $Z_3$ symmetry generated by $ST$ is unbroken. Small deviation $\delta \tau ={\cal O}(0.01)$ would be 
phenomenologically interesting \cite{Petcov:2022fjf,Kikuchi:2023jap}.

The modular forms of weight $k_Y$ and level $N$ transform under 
$ST \in \Gamma_N$ 
as,
\begin{equation}
\label{eq: ST-trans.}
 [Y_{\bm{r}}^{(k_Y)}(\tau)]_i \xrightarrow{ST}   [Y_{\bm{r}}^{(k_Y)}(-1/(\tau+1))]_i = (-1-\tau)^{k_Y} \rho_{\bm{r}}(ST)_{ij}  [Y_{\bm{r}}^{(k_Y)}(\tau)]_j,
\end{equation}
where ${\bm r}$ denotes the representation. Since we will only focus on singlet representations, indices $i,j$ are not relevant hereafter. The representation matrix corresponding to a singlet is written as $\rho_{\bm{r}}(ST) = \omega^{q_{\bm{r}}}$ where $q_{\bm{r}} \in \mathbb{Z}$ denotes the $ST$-charge of the modular form $Y_{\bm{r}}^{(k_Y)}$.
For convenience, we introduce the following complex variable 
\begin{equation}
    u \equiv \frac{\tau - \omega}{\tau - \omega^2},
\end{equation}
which parametrizes the deviation of $\tau$ from the $ST$-invariant fixed point $\omega$. Notice that $u$ is transformed to $\omega^2 u$ under the $ST$.
We can rewrite Eq.\,(\ref{eq: ST-trans.}) as
\begin{equation}
\label{eq: ST-trans_2}
    Y_{\bm{r}}^{(k_Y)}(\omega^2 u) = \left( \frac{1-\omega^2u}{1-u} \right)^{k_Y} \tilde{\rho}_{\bm{r}}(ST)  Y_{\bm{r}}^{(k_Y)}(u),
\end{equation}
where $\tilde{\rho}_{\bm{r}}= \omega^{-k_Y} \rho_{\bm{r}}$. Since both sides of Eq.\,(\ref{eq: ST-trans_2}) are holomorphic with respect to $u$, we expand both sides in $u$, which yields
\begin{equation}
\label{eq: u-expansion}
\omega^{2l}
\frac{d^l\tilde{Y}_{\bm{r}}^{(k_Y)}(u)}{du^l}\bigg|_{u=0}
= \tilde{\rho}_{\bm{r}}(ST) \frac{d^l \tilde{Y}_{\bm{r}}^{(k_Y)}(u)}{du^l}\bigg|_{u=0}
,\quad (l=0,1,2,\cdots),
\end{equation}
where $\tilde{Y}_{\bm{r}}^{(k_Y)} = (1-u)^{-k_Y} Y_{\bm{r}}^{(k_Y)}$. We obtain
\begin{equation}
\label{eq: ST-invariance}
    (\omega^{2l} - \omega^{q_{{\bm{r}}}-k_Y})
    \frac{d^l \tilde{Y}_{\bm{r}}^{(k_Y)}(u)}{du^l}\bigg|_{u=0}
    =0.
\end{equation}
This shows that $
\frac{d^l \tilde{Y}_{\bm{r}}^{(k_Y)}(u)}{du^l}\big|_{u=0} = 0
$ unless $2l \equiv q_{{\bm{r}}} - k_Y \pmod{3}$. 
Thus, we expect that the modular forms become hierarchical depending on their $ST$-charges when $\tau$ is close to $\omega$.  Note also that
\begin{equation}
     Y_{\bm{r}}^{(k_Y)}(\tau ) = (1-u)^{k_Y} \tilde{Y}_{\bm{r}}^{(k_Y)},
\end{equation}

\begin{equation}
\label{eq: dY/dtau}
    \frac{d Y_{\bm{r}}^{(k_Y)}(\tau)}{d\tau} = \frac{(1-u)^{k_Y+2}}{\sqrt{3}i} \left[ \frac{d\tilde{Y}_{\bm{r}}^{(k_Y)}}{du} - \frac{k_Y}{1-u} \tilde{Y}_{\bm{r}}^{(k_Y)} \right],
\end{equation}

\begin{align}
\begin{aligned}
\label{eq: d^2Y/dtau^2}
    \frac{d^2 Y_{\bm{r}}^{(k_Y)}(\tau)}{d\tau^2}
     = - \frac{(1-u)^{k_Y+4}}{3} \left[
     \frac{d^2 \tilde{Y}_{\bm{r}}^{(k_Y)}}{du^2} 
    -2\frac{k_Y + 1}{1-u}  \frac{d\tilde{Y}_{\bm{r}}^{(k_Y)}}{du} + \frac{k_{Y}^2+k_Y}{(1-u)^2} \tilde{Y}_{\bm{r}}^{(k_Y)} \right],
\end{aligned}
\end{align}
which will be used in the later discussion.

\section{Moduli stabilization in $A_4$ model}
\label{sec:model}
For concreteness, we study the stabilization of the complex structure modulus $\tau$ in $A_4 \simeq \Gamma_3$ modular flavor symmetric models 
and discuss the 1-loop effective potential $V_{\rm eff}(\tau,\bar{\tau})$ within the models.

In order to proceed further, we assume the following superpotential
\begin{equation}
\label{eq: A4_superpotential}
    W(\tau, \Phi_I) = \frac{1}{2} \langle \phi \rangle Y_{\bm r}^{(8)}(\tau) \sum_{I=1}^n \Phi_I^2,
\end{equation}
where $\langle \phi \rangle$ is a mass parameter.\footnote{One of the possible origins of the superpotential in Eq.\,(\ref{eq: A4_superpotential}) 
is the D-brane instanton effect. 
In such a case, $\langle \phi \rangle$ corresponds to $\sim e^{-S_{\rm cl}}M_{\rm com}$ where $S_{\rm cl}$ is the classical action of the D-brane instanton and $M_{\rm com}$ denotes the compactification scale\cite{Ibanez:2012zz}. We may also regard $\langle \phi\rangle$ as the VEV of a GUT Higgs superfield.} $Y_{\bm{r}}^{(8)}(\tau)$ denotes the modular form of level $3$ with modular weight $k_Y=8$ which belongs to the representation $\bm{r}$. For simplicity, we will treat modular forms in the singlet representations $\bm{r} = \bm{1}, \bm{1}'$, and $\bm{1}''$ of $A_4 \simeq \Gamma_3$. 
Those $A_4$ modular forms are explicitly defined in Appendix \ref{sec: A_4} and corresponding representation matrices are summarized in Appendix \ref{appendix: group_A4}.
Note that $k_Y=8$ is the lowest modular weight where there exist three non-vanishing $A_4$ singlet modular forms of $\bm{1}, \bm{1}'$ and $\bm{1}''$.
For later convenience, we show the behavior of the $A_4$ modular forms in the vicinity of the fixed point $\tau = \omega$,
\begin{align}
\label{eq: Y1_series}
Y_{\bm{1}}^{(8)}(\tau) &= -\frac{1}{6} 
\frac{d^2 \tilde{Y}^{(8)}_{\bm{1}}(u)}{du^2}\bigg|_{u = 0}
(\tau - \omega)^2 + \mathcal{O}((\tau - \omega)^3), \\
\label{eq: Y1'_series}
Y_{\bm{1}'}^{(8)}(\tau) &= \frac{1}{\sqrt{3}i} 
\frac{d \tilde{Y}^{(8)}_{\bm{1}'}(u)}{du}\bigg|_{u = 0}
(\tau - \omega) + \mathcal{O}((\tau - \omega)^2), \\
\label{eq: Y1''_series}
Y_{\bm{1}''}^{(8)}(\tau) &= \tilde{Y}^{(8)}_{\bm{1}''}(0) -\frac{8\tilde{Y}^{(8)}_{\bm{1}''}(0)}{\sqrt{3}i}  (\tau - \omega) + \mathcal{O}((\tau - \omega)^2).
\end{align}
We assume that the matter chiral superfields $\Phi_I$ also belong to one of the three $A_4$ singlet representations with an integral weight $-k_I$.
For the modular invariance of the superpotential, $k_I=4, (\forall I)$ is required. 
If $k_I \neq 4$, it implies that $\langle \phi \rangle$ has a non-zero modular weight, hence modular symmetry is broken.\footnote{Modular $T$-symmetry can remain unbroken even when the modular weights do not cancel between modular forms and chiral superfields.} 

We note that unless SUSY is spontaneously broken, the one-loop effective potential identically vanishes, but in realistic models SUSY must be broken at some scale. In order to keep generality of our discussion, we will introduce soft SUSY breaking terms without specifying their origins, with which there appear non-vanishing one-loop effective potential denoted by $V_1$. In particular, in the presence of $n$-different matter flavors, $V_1$ would be multiplied by the flavor number $n$.
Thus, the effective potential in the one-loop approximation is given by
\begin{equation}
\label{eq: V_eff}
    V_{\rm eff} = V_0 + n V_1,
\end{equation}
where $n$ denotes the number of chiral superfields and $V_0$ corresponds to the tree-level potential.

Let us comment on the tree level potential $V_0$. It was shown that the fixed
point $\tau = \omega$ is favored statistically with the highest probability \cite{Ishiguro:2020tmo} within the statistics of three form flux superpotential in string theory. Thus, it is reasonable to assume that the tree level potential stabilizes $\tau$ near the fixed point $\omega$.
For simplicity of our following analysis, we will approximate the potential as 
\begin{equation}\label{V0form}
    V_0 = m_{\tau}^4 |\tau - \omega|^2,
\end{equation}
where $m_{\tau}$ is a mass parameter. We expect that the details of the tree level potential lost in our approximation would not change our conclusion as long as the deviation from the minimum $\tau=\omega$ is small enough.

\subsection{Coleman-Weinberg potential}
The Coleman-Weinberg potential  $V_1$ is straightforwardly computed as
\begin{equation}
\label{eq: V_CW}
V_{1}
= \frac{1}{32 \pi^2} \Big[(M^2+m_0^2)^2  \log \left( \frac{ M^2+m_0^2}{\sqrt{e} \Lambda^2} \right) - M^4  \log \left( \frac{ M^2}{\sqrt{e} \Lambda^2} \right) \Big],
\end{equation}
at the one-loop level,\footnote{Terms that vanish when $\Lambda^2$ goes to infinity are neglected~\cite{Coleman:1973jx}.
Thus, we need the condition $M^2 + m_0^2 \ll \Lambda^2$ to trust the expression in Eq.\,(\ref{eq: V_CW}) as a good approximation. In particular, we require $\max(M^2,m_0^2)/\Lambda^2 \lessapprox 0.01$ for the validity of our approximation.} 
where $M^2$ is defined as 
\begin{equation}
\label{eq: M^2}
    M^2 = (2{\rm Im}\tau)^{2k_I} \langle \phi \rangle^2 \left| Y^{(k_Y)}_{\bm{r}} \right|^2,
\end{equation}
and $\Lambda$ denotes the cut-off 
which we will take to be near the compactification scale.\footnote{More precisely, we have used the cut-off regularization and identified the cut-off scale to be the compactification scale. Even if we use the dimensional regularization, we would find the same result by imposing appropriate renormalization conditions.} The factor $(2{\rm Im}\tau)^{2k_I}$ in Eq.\,(\ref{eq: M^2}) comes from the redefinition of component fields to normalize their kinetic terms into canonical ones. The first term on the right-hand side of Eq.\,(\ref{eq: V_CW}) corresponds to the bosonic contributions, while the second term corresponds to those of fermions. 
We have introduced the soft SUSY breaking mass $m_0$ to the scalar components by hand. In particular, we have assumed that $m_0$ is independent of $\tau$.\footnote{Such a situation may be realized if the SUSY breaking is mediated by a field that does not couple to $\tau$, particularly its modular weight should be zero. See for modular symmetry of soft SUSY breaking terms Ref.~\cite{Kikuchi:2022pkd}.}
Notice that for $k_I=4$ where $\langle \phi \rangle$ is a modular singlet, $M^2$ and accordingly $V_1$ are modular invariant as they should be. Moreover, $V_1$ is invariant under the CP, $\tau \rightarrow -\bar{\tau}$. 

For our purpose, we summarize the derivatives of the potential as follows: The first derivatives of $V_1$ with respect to $x \equiv {\rm Re} \tau$ and $y \equiv {\rm Im}\tau$ are given by 
\begin{align}
\begin{aligned}
\label{eq: dV}
\frac{\partial V_{1}}{\partial x} &= \frac{1}{32\pi^2} \frac{\partial M^2}{\partial x} \  C(M^2,m_0^2), \\
\frac{\partial V_{1}}{\partial y} &= \frac{1}{32\pi^2} 
\frac{\partial M^2}{\partial y} \ C(M^2,m_0^2),
\end{aligned}
\end{align}
where
\begin{equation}
\label{eq: Curve_C}
    C(M^2,m_0^2) \equiv  2(M^2 + {m_0}^2)  \log \Big( \frac{ M^2+m_0^2}{\sqrt{e} \Lambda^2} \Big) - 2 M^2 \log \left( \frac{ M^2}{\sqrt{e} \Lambda^2} \right) + {m_0}^2.
\end{equation}
The second derivatives are given by 
\begin{align}
\begin{aligned}
\label{eq: ddV}
\frac{\partial^2 V_{1}}{\partial x^2} 
&= \frac{1}{32\pi^2}  \left[ \frac{\partial^2 M^2}{\partial x^2} \ C(M^2,m_0^2) +2\left( \frac{\partial M^2}{\partial x} \right)^2 \log{\left( \frac{M^2+m_0^2}{M^2} \right)} \right],\\
\frac{\partial^2 V_{1}}{\partial y^2} 
&= \frac{1}{32\pi^2}  \left[ \frac{\partial^2 M^2}{\partial y^2} \ C(M^2,m_0^2) +2\left( \frac{\partial M^2}{\partial y} \right)^2 \log{\left( \frac{M^2+m_0^2}{M^2} \right)} \right],\\
\frac{\partial^2 V_{1}}{\partial x \partial y} 
&= \frac{1}{32\pi^2} \left[ \frac{\partial^2 M^2}{\partial x \partial y}\ C(M^2,m_0^2) +2\left( \frac{\partial M^2}{\partial x} \right)  \left( \frac{\partial M^2}{\partial y} \right) \log{\left( \frac{M^2+m_0^2}{M^2} \right)} \right],
\end{aligned}
\end{align}
where 
\begin{align}
\begin{aligned}
    \frac{\partial M^2}{\partial{x}} &= 2{\langle \phi \rangle}^2 (2y)^{2k_I}\ {\rm Re}\{(\partial_{\tau}Y)Y^{*}\}, \\
     \frac{\partial M^2}{\partial{y}} &=  2^{2k_I} {\langle \phi \rangle}^2 y^{2k_I-1} \left[ -2y\ {\rm Im}\{(\partial_{\tau}Y)Y^{*}\} +2k_I |Y|^2 \right],
\end{aligned}
\end{align}
and 
\begin{align}
\begin{aligned}
\frac{\partial^2 M^2}{\partial{x}^2} &= 2 {\langle \phi \rangle}^2 (2y)^{2k_I} \left[ {\rm Re}\{ (\partial_{\tau}^2 Y)Y^* \} + |\partial_{\tau}Y|^2  \right],\\
\label{eq: DM/Dy}
\frac{\partial^2 M^2}{\partial{y}^2} &= 2^{2k_I}  {\langle \phi \rangle}^2 y^{2k_I-2} \left[-2y^2 
( {\rm Re}\{ (\partial_{\tau}^2 Y)Y^* \} - |\partial_{\tau}Y|^2) -8k_Iy{\rm Im}\{(\partial_{\tau}Y)Y^* \} +2k_I(2k_I-1)|Y|^2 \right], \\
\frac{\partial^2 M^2}{\partial{x} \partial{y}} &= 2^{2k_I+1} {\langle \phi \rangle}^2 y^{2k_I-1} \left[2k_I {\rm Re}\{ (\partial_{\tau} Y)Y^* \} - y {\rm Im} \{(\partial_{\tau}^2 Y)Y^*  \} \right].
\end{aligned}
\end{align}

\subsection{Canonical basis of modulus}
The kinetic term of the modulus field $\tau$ resulting from Eq.\,(\ref{eq: Kahler}) is given by 
\begin{equation}
\label{eq: kinetic_term_tau}
  - \frac{\Lambda_0^2}{(2{\rm Im}\tau)^2} |\partial_{\mu} \tau|^2 = - \frac{\Lambda_{0}^2}{(2 y)^2} \left[ (\partial_{\mu} x)^2 +(\partial_{\mu} y)^2  \right],
\end{equation}
where $x={\rm Re}\tau$ and $y={\rm Im} \tau$ which are not yet canonically normalized. In order to discuss the stability of the vacua from the potential analysis, we need to convert the basis to canonical ones. However, the second derivatives of the potential evaluated at stationary points do not change signs under the conversion. This means that a local minimum (maximum) of a potential $V$ plotted as a function of $x$ and $y$ is still a local minimum (maximum) of the ones in the canonical basis.

One may confirm the above statement as follows: Consider a stationary point of a potential $V(x,y)$ given by $\tau_* = x_* + iy_*$. In the vicinity of the stationary point, we may expand the kinetic term in Eq.\,(\ref{eq: kinetic_term_tau}) as,
\begin{align}
\label{eq: non-canonical}
    - \Lambda_0^2 \left[\frac{1}{(2y_*)^2} + \mathcal{O}(\mathit{\Delta} y) \right] \cdot \left[ (\partial_{\mu} \mathit{\Delta} x)^2 +(\partial_{\mu} \mathit{\Delta} y)^2  \right],
\end{align}
where $\mathit{\Delta}x = x-x_*$ and $\mathit{\Delta} y = y - y_*$.
On the other hand, in terms of canonical basis $\chi, \psi$ we have
\begin{equation}
\label{eq: canonical}
\mathcal{L}_{\rm kin}= - \frac{1}{2} \left[ (\partial_{\mu} \chi )^2 +(\partial_{\mu} \psi)^2  \right].
\end{equation}
By comparing Eq.\,(\ref{eq: non-canonical}) and Eq.\,(\ref{eq: canonical}), we obtain
\begin{align}
\begin{aligned}
\label{eq: linear_canonical}
     \chi(\mathit{\Delta} x, \mathit{\Delta} y) &= \frac{\Lambda_0}{\sqrt{2}y_*}  \mathit{\Delta} x + \mathcal{O}(\mathit{\Delta}^2),\\
 \psi(\mathit{\Delta} x, \mathit{\Delta} y) &= \frac{\Lambda_0}{\sqrt{2}y_*}  \mathit{\Delta} y + \mathcal{O}(\mathit{\Delta}^2),
\end{aligned}
\end{align}
where $\mathcal{O}(\mathit{\Delta}^2)$ denotes second or higher order terms of $\mathit{\Delta} x$ and $\mathit{\Delta} y$. Then we find
\begin{align}
\begin{aligned}
\frac{\partial^2 V}{\partial \chi^2} \bigg|_{\tau = \tau_*} 
&= \frac{2y_*^2}{\Lambda_0^2} \frac{\partial^2 V}{\partial x^2} \bigg|_{\tau = \tau_*} ,\\
\frac{\partial^2 V}{\partial \psi^2} \bigg|_{\tau = \tau_*} 
&= \frac{2y_*^2}{\Lambda_0^2} \frac{\partial^2 V}{\partial  y^2} \bigg|_{\tau = \tau_*} ,\quad   \\
\frac{\partial^2 V}{\partial \chi \partial \psi} \bigg|_{\tau = \tau_*} &= 
 \frac{2y_*^2}{\Lambda_0^2} \frac{\partial^2 V}{\partial  x \partial y} \bigg|_{\tau = \tau_*}.
\end{aligned}
\end{align}
Thus, the second derivatives of potential around the minimum/maximum are the same up to an overall factor, and therefore, one may simply discuss the stability with non-canonical variables.

\section{Vacuum structure with radiative corrections}
\label{sec:results}

In this section, we analyze the behavior of the 1-loop Coleman-Weinberg potential $V_1$
especially around the fixed point $\tau = \omega$ for each model with the $A_4$ modular form of $\bm{r}=\bm{1}, \bm{1}', \bm{1}''$.  We also discuss the possibility to obtain stable vacua in the vicinity of $\omega$ by considering the effective potential $V_{\rm eff}=V_0 + n V_1$.

\subsection{$\bf{1}$}
Here, we study the case when the $A_4$ modular form of representation $\bm{r}=\bm{1}$ with weight $k_Y=8$, namely, we take $Y_{\bm{1}}^{(8)}$ in Eq.\,(\ref{eq: M^2}).
In the vicinity of $\tau = \omega$, we have 
\begin{equation}
\label{eq: M^2_1_expand}
    M^2 = \frac{3^{k_I} \langle \phi \rangle^2 }{36} 
\bigg| \frac{d^2\tilde{Y}_{\bf{1}}^{(8)}(u)}{du^2}\bigg|_{u = 0} \bigg|^2 
|\tau - \omega|^4 + \cdots,
\end{equation}
from Eq.\,(\ref{eq: Y1_series}).\footnote{In the vicinity of $\tau=\omega$, we find $M^2 \ll 1$. Thus, the Coleman-Weinberg potential in Eq.\,(\ref{eq: V_CW}) is valid in this region.}
It follows from Eqs.\,(\ref{eq: dV}) and (\ref{eq: ddV}) that both the first and second derivatives of the 1-loop quantum correction $V_1$ with respect to $x={\rm Re}\tau$ and $ y={\rm Im}\tau$ vanish at the fixed point,
\begin{align}
\begin{aligned}
    \frac{\partial V_1}{\partial x}\bigg|_{\tau = \omega}  &=
    \frac{\partial V_1}{\partial y}\bigg|_{\tau = \omega}
    = 0,\\
    \frac{\partial^2 V_1}{\partial x^2}\bigg|_{\tau = \omega} 
    &= 
    \frac{\partial^2 V_1}
    {\partial y^2}\bigg|_{\tau = \omega} = 
    \frac{\partial^2 V_1}{\partial x \partial y}\bigg|_{\tau= \omega} = 0,
\end{aligned}
\end{align}
 while $V_1(\omega)= \frac{1}{32\pi^2} m_0^4 \log{ \left( \frac{m_0^2}{\sqrt{e}\Lambda^2} \right)}$.

We have shown a numerical illustration of $V_1$ in Fig.~\ref{fig:V1,phi=0.01,k=8,near_omega}, which clearly shows the instability of the point $\tau=\omega$. Indeed, the fixed point $\tau = \omega$ is a local maximum of $V_1$ independently of the values of $k_I, \langle \phi \rangle$ and $m_0$.
We can understand such a behavior by expanding $V_1$ in $\frac{M^2}{m_0^2}$, which is a good approximation since $M^2 \rightarrow 0$ as $\tau \rightarrow \omega$.
If $\tau$ is sufficiently close to $\omega$, the bosonic contribution dominates the potential $V_1$. We may approximate it as   
\begin{equation}
    V_1 \simeq \frac{m_0^4}{32\pi^2} \left[  \log{\left( \frac{m_0^2}{\sqrt{e}\Lambda^2} \right)} + \frac{2M^2}{m_0^2} \log{\left( \frac{m_0^2}{\Lambda^2} \right)} \right].
\end{equation}
Taking account of  $\log(m_0^2/\Lambda^2)<0$ and Eq.(\ref{eq: M^2_1_expand}), the second term becomes an inverted quartic potential maximized at $\tau=\omega$. 
Thus, the combination of the tree level potential $V_0 = m_{\tau}^4 |\tau - \omega|^2$ and the one-loop correction $nV_1$ would hardly realize the desired small deviation from the fixed point $\tau=\omega$.

\begin{figure}[H]
\centering
\includegraphics[width=100mm]{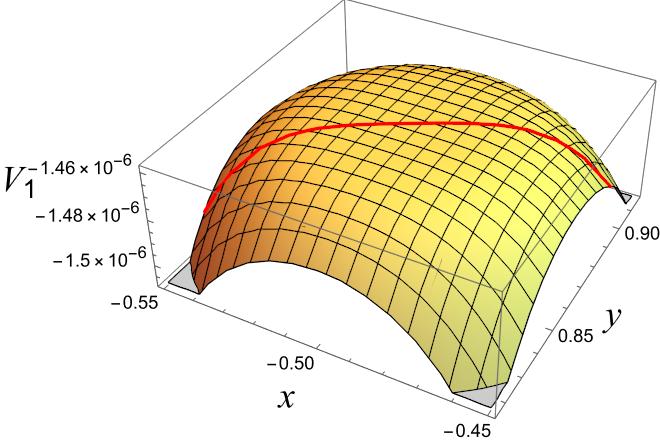}
\caption{
The one-loop potential $V_1$ with ${\bm r}={\bm 1}$ and $k_Y=8$
in the vicinity of $\tau = \omega$.
The parameters are chosen as $k_I=4,\langle \phi \rangle=10^{-2},m_0=10^{-1}$ under $\sqrt{e}\Lambda^2 = 1$. The red line shows the arc $|\tau|=1$.}
\label{fig:V1,phi=0.01,k=8,near_omega}
\end{figure}

We briefly comment on the behavior of the one-loop potential far from the fixed point $\tau=\omega$. From Eq.\,(\ref{eq: dV}), one may think that there can be additional minima at the point satisfying $C(M^2,m_0^2)=0$. However, such points are generally unphysical as they are inconsistent with our effective field theory description valid only if $M^2\ll \Lambda^2$.

\subsection{$\bf{1}'$}
We proceed to the case where the $A_4$ modular form of representation $\bm{r}=\bm{1}'$ with weight $k_Y=8$, namely $Y_{\bm{1}'}^{(8)}$ in Eq.\,(\ref{eq: M^2}). 
In the vicinity of $\tau = \omega$, we have 
\begin{equation}
\label{eq: M^2_1'_expand}
    M^2 = \frac{3^{k_I} |D|^2 \langle \phi \rangle^2}{3} |\tau - \omega|^2 \left[ 1 + \delta y \left( \frac{4k_I}{\sqrt{3}} - 6 \sqrt{3} \right) + \mathcal{O}(|\tau - \omega|^2)
    \right],
\end{equation}
from Eq.\,(\ref{eq: Y1'_series}) where $D=
\frac{d \tilde{Y}^{(8)}_{\bf{1}'}(u)}{du} \big|_{u=0} \simeq -10.6+18.3i$ and we have defined $\delta y = y- \frac{\sqrt{3}}{2}$.
It follows that the first derivatives of the 1-loop quantum correction $V_1$ with respect to $x={\rm Re}\tau$ and $y={\rm Im}\tau$ vanish at the fixed point,
\begin{align}
    \frac{\partial V_1}{\partial x}\bigg|_{\tau = \omega}  &=
    \frac{\partial V_1}{\partial y}\bigg|_{\tau = \omega} = 0,
\end{align}
whereas the second derivatives are non-vanishing,
\begin{align}
    \frac{\partial^2 V_1}{\partial x^2}\bigg|_{\tau= \omega}
 &= 
    \frac{\partial^2 V_1}{\partial y^2}\bigg|_{\tau =\omega} = \frac{1}{24 \pi^2} 3^{k_I} \left| D \right|^2 \langle \phi \rangle^2  m_0^2 \log{\left( \frac{m_0^2}{\Lambda^2} \right)}, \\
    \frac{\partial^2 V_1}{\partial x \partial y}\bigg|_{\tau= \omega} &= 0.
\end{align}
We find 
$ \frac{\partial^2 V_1}{\partial x^2}\big|_{\tau=\omega} < 0$ and $\frac{\partial^2 V_1}{\partial y^2}\big|_{\tau=\omega} < 0$, hence the fixed point $\tau = \omega$ is a local maximum of $V_1$ independent of the values of $k_I, \langle \phi \rangle$ and $m_0^2$.\footnote{Note that $m_0^2 \ll \Lambda^2$.} Figure\,\ref{tb:V2,phi=0.01,k=8,near_omega} shows $V_1$ in the vicinity of $\tau = \omega$ with the parameters $k_I=4, \langle \phi \rangle=10^{-2}, m_0=10^{-1}$. 
\begin{figure}[htbp]
\centering
\includegraphics[width=100mm]{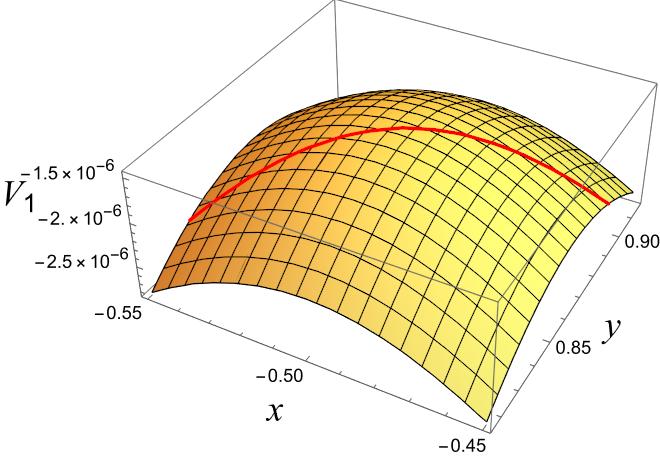}
\caption{The one-loop effective potential $V_1$
in the vicinity of $\tau = \omega$ for $\bm{r}=\bm{1}'$ with the weight $k_Y=8$. The parameters are chosen as $k_I=4,\langle \phi \rangle=10^{-2},m_0=10^{-1}$ under   $\sqrt{e}\Lambda^2 = 1$. The red line shows the arc $|\tau|=1$.}
\label{tb:V2,phi=0.01,k=8,near_omega}
\end{figure}

Unlike the previous case with $\bm r= \bm{1}$, the second derivatives are non-vanishing and therefore there is a possibility to realize a small deviation from the fixed point. Nevertheless, the one-loop effective potential is generally small, and one of ways to realize the deviation is to increase the number of species contributing the loop correction.
We are able to estimate the number of species $n$ that makes the one-loop correction to be compatible with the tree level potential as
 \begin{equation}
 \label{eq: n_c}
n \geq  n_c \equiv \frac{-48\pi^2 m_{\tau}^4}{3^{k_I} |D|^2 \langle \phi \rangle^2  m_0^2 \log{(m_0^2/\Lambda^2)}} ,
\end{equation}
where $n_c$ satisfies
\begin{equation}
 \frac{\partial^2}{\partial x^2}(V_0 + n_c V_1) =0,\   \frac{\partial^2}{\partial y^2}(V_0 + n_c V_1)=0,\quad ({\rm at} \ \tau = \omega).
\end{equation}
Eq.\,(\ref{eq: n_c}) shows that larger the value of $m_0$ and $\langle \phi \rangle$, smaller the value of $n_c$.

We emphasize that the compatibility of the ``tree-level potential'' and the one-loop potential does not imply the breakdown of our perturbative approach. This is because the origin of the tree-level potential we assume here has generally nothing to do with the one-loop contribution. For instance, the ``tree-level potential'' can originate from non-perturbative effects, such as gaugino condensation. On the other hand, the Coleman-Weinberg potential originates from the loops of matter fields coupling to $\tau$ through their ``Yukawa couplings''. Therefore, the ``tree-level potential'' and the one-loop corrections can be compatible without problems of strong couplings. 
Indeed, the coupling is small $|Y(\tau)| \leq \mathcal{O}(0.1)$ when the modulus is in the vicinity of $\tau = \omega$. 
Nevertheless, increasing the number of species may conflict with the perturbativity of gravitational interactions~\cite{Dvali:2007wp}, known as the ``species bound''. In our case, thanks to SUSY, we expect that the one-loop correction to the Newton constant is relaxed. Nevertheless, a naive application of the species bound requires
\begin{align}
\label{eq: species_bound}
    \Lambda<\frac{M_{\rm pl}}{\sqrt{n}},
\end{align}
where $M_{\rm pl}$ is the Planck scale. If 
we take $n\sim 10^3$-$10^7$ as shown in our numerical examples later, 
that yields roughly $\Lambda<10^{14}$-$10^{16}{\rm GeV}$. Therefore, the cut-off scale ($\sim$ the compactification scale) needs to be below such a scale, which can be satisfied without any problems. 
On the other hand, if we take $M_{\rm pl}\simeq 10^{19}$GeV and the cut-off $\Lambda$ 
near the compactification scale $M_{\rm com}\simeq 10^{17}$GeV. Then, the condition Eq.\,(\ref{eq: species_bound}) requires $n < 10^4$. In our numerical examples shown later, we will see that this ``species bound" is satisfied if $m_{\tau}$ is reasonably small.

We comment on a property of the effective potential. As a consequence of the CP-invariance $\tau\to-\bar\tau$, the effective potential is invariant under $\delta x\to-\delta x$, where $\delta x=x+\frac12$. 
Note that spontaneous CP violation\footnote{CP-violation is caused by the VEV of $\tau$ if it lies neither on the lines $2x \equiv 0 \pmod{1}$ nor the arc $|\tau|=1$. } would have important phenomenological impacts particularly on the flavor structure. However, as will be shown in Eq.\,(\ref{eq: approx_V1}), such violation would not take place 
within our model near $\tau = \omega$. 
Thus, our primary interest is in the behavior of the effective potential along the line $x=-\frac12$.

\begin{figure}[H]
\centering
  \begin{minipage}[b]{0.45\linewidth}
    \centering
 \includegraphics[width=70mm]{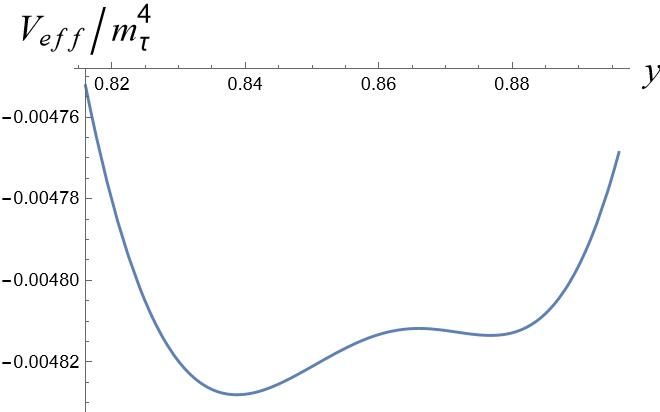}
\caption{$n/m_{\tau}^4=3300$}
\label{fig: V2_k8_phi0.01_m0.1_n_3300_section}
  \end{minipage}
   \begin{minipage}[b]{0.45\linewidth}
    \centering
\includegraphics[width=70mm]{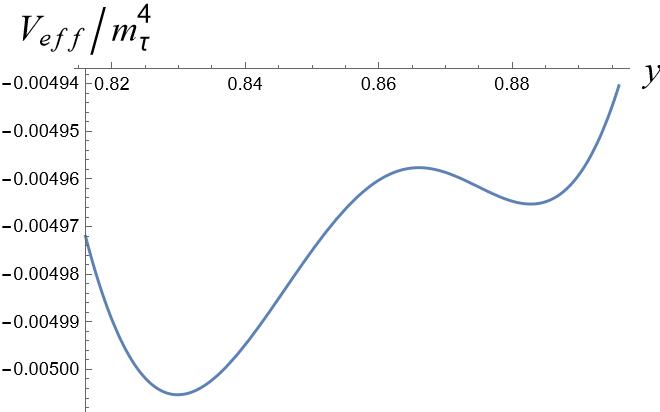}
\caption{$n/m_{\tau}^4=3400$}
\label{fig: V2_k8_phi0.01_m0.1_n_3400_section}
  \end{minipage}
\end{figure}

We have shown $V_{\rm eff}$ along $x=-\frac12$ with $(k_I, \langle \phi \rangle, m_0)=(4, 10^{-2}, 10^{-1})$ and $n/m_{\tau}^4 = 3300$ (Fig.~\ref{fig: V2_k8_phi0.01_m0.1_n_3300_section}) or $n/m_{\tau}^4 = 3400$ (Fig.~\ref{fig: V2_k8_phi0.01_m0.1_n_3400_section}).
For those cases, we have numerically derived the deviation $|\delta y_*|\simeq 0.0273$ and $|\delta y_*| \simeq 0.0362$, respectively, where $\delta y_* = y_* - \frac{\sqrt{3}}{2}$.
Note that we have found the critical species number to be $n_c/m_{\tau}^4 \simeq 3180$ in the unit 
satisfying 
$\sqrt{e} \Lambda^2 = 1$,
from which we have chosen the parameters above.
For clarity, we present explicit values of parameters with $e^{1/4} \Lambda = 10^{17}$GeV. 
When $m_{\tau}$ is $10^{17}$GeV, $0.5 \times 10^{17}$GeV, $0.2 \times 10^{17}$GeV, the corresponding critical species number is given by $n_{c} \simeq 3180, 200, 5$, respectively. These examples are consistent with the species bound~\eqref{eq: species_bound}. 

\begin{figure}[htbp]
\centering
\includegraphics[width=82mm]{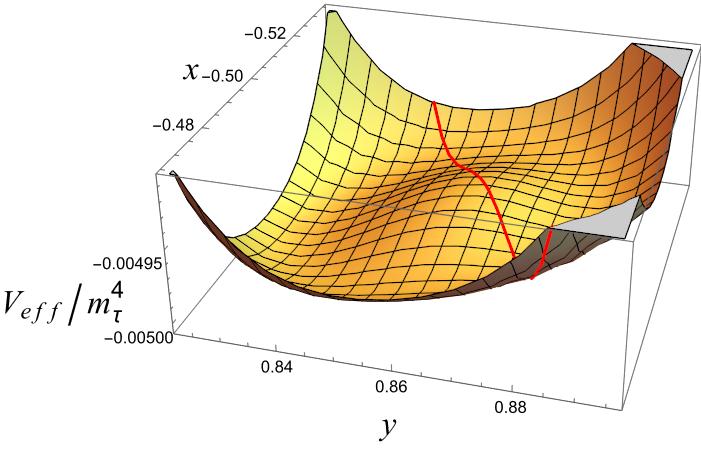}
\caption{Three-dimensional plot of $V_{\rm eff}/m_{\tau}^4$
in the vicinity of $\tau = \omega$.
The chosen parameters are identical to those in Fig.~\ref{fig: V2_k8_phi0.01_m0.1_n_3400_section}, namely $k_I=4,\langle \phi \rangle=10^{-2},m_0=10^{-1}, n/m_{\tau}^4=3400$ in the unit $\sqrt{e}\Lambda^2 = 1$.
The red line shows the arc $|\tau|=1$.}
\label{fig: V2_k8_phi0.01_m0.1_n_3400}
\end{figure}

\begin{figure}[H]
\centering
\begin{minipage}[b]{0.45\linewidth}
    \centering
 \includegraphics[width=70mm]{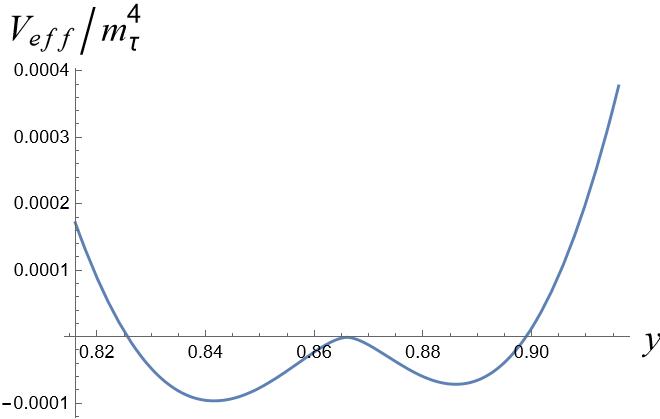}
\caption{$n/m_{\tau}^4=2.2 \times 10^7$}
\label{fig: V2_k8_2.2*10^7}
  \end{minipage}
   \begin{minipage}[b]{0.45\linewidth}
    \centering
\includegraphics[width=70mm]{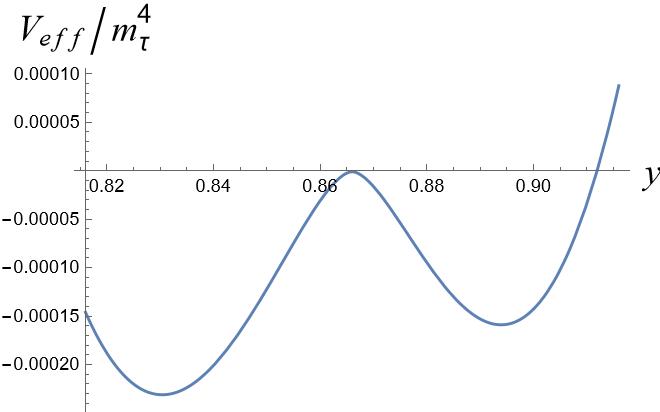}
\caption{$n/m_{\tau}^4=2.5 \times 10^7$}
\label{fig: V2_k8_2.5*10^7}
  \end{minipage}
\end{figure}

\begin{figure}[H]
\centering
\includegraphics[width=95mm]{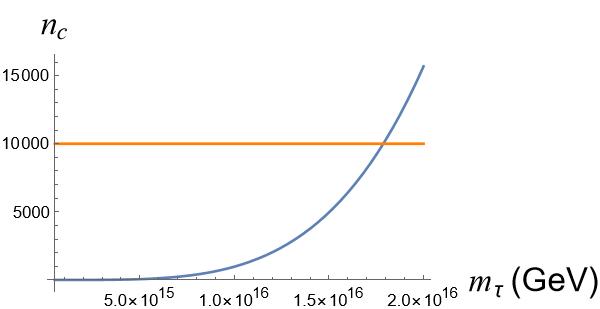}
\caption{The relation between $n_c$ and $m_\tau$. We have shown $n_c/(m_{\tau})^4=9.8 \times 10^6$ (blue curve) and also the upper bound on the species number $(\Lambda^2/M_{\rm pl})^2 \simeq 10^4$ (orange line). We have used $e^{1/4} \Lambda =10^{17}$GeV.}
\label{fig: n/m^4=9.8*10^6}
\end{figure}

In confirmation of the vacuum stability, we also show the $(x,y)$ dependence of $V_{\rm eff}$ in Fig.~\ref{fig: V2_k8_phi0.01_m0.1_n_3400} where we have used the same parameters as in Fig.~\ref{fig: V2_k8_phi0.01_m0.1_n_3400_section}. We have confirmed that the local minimum is on the ``CP invariant'' line $\delta x=0$. The shape of 1-loop corrected potential generally takes the ``tilted wine-bottle'' shape as the one presented in Fig.~\ref{fig: V2_k8_phi0.01_m0.1_n_3400}.  
We also show the case of a different parameter set, $(k_I, \langle \phi \rangle, m_0)=(4, 10^{-2}, 10^{-3})$ with $n/m_{\tau}^4 = 2.2 \times 10^7$ (Fig.~\ref{fig: V2_k8_2.2*10^7}) or $n/m_{\tau}^4 = 2.5 \times 10^7$ (Fig.~\ref{fig: V2_k8_2.5*10^7}). 
We have numerically derived the deviation $|\delta y_*|\simeq 0.0244$ and $|\delta y_*| \simeq 0.0356$, respectively. 
In this case, we have found $n_c/m_{\tau}^4 \simeq 9.8 \times 10^6$. As numerical examples, the critical species numbers are $n_{c} \simeq 9.8 \times 10^6, 980, 61$ for $m_\tau=10^{17}$GeV, $10^{16}$GeV, $0.5 \times 10^{16}$GeV, respectively, where we have taken $e^{1/4} \Lambda = 10^{17}$GeV.
In Fig.~\ref{fig: n/m^4=9.8*10^6}, we have shown the relation between $m_{\tau}$ and $n_c$ with the ``species bound".
The blue line represents the relation $n_c/m_{\tau}^4=9.8 \times 10^6$ while the orange line corresponds to the upper bound of the species number $n \leq (\Lambda^2/M_{\rm pl})^2 \simeq 10^4$. Thus, our model is within the bound if $m_{\tau}$ is reasonably small (e.g. $m_{\tau} \lessapprox 1.5 \times 10^{16}$GeV).

\begin{figure}[H]
\centering
\includegraphics[width=100mm]{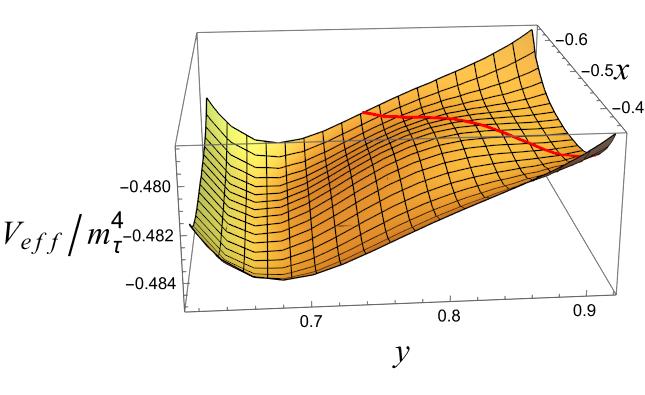}
\caption{Three-dimensional plot of $V_{\rm eff}/m_{\tau}^4$.
The parameters are chosen as $k_I=4,\langle \phi \rangle=10^{-3},m_0=10^{-1}, n/m_{\tau}^4=3.3 \times 10^5$ under   $\sqrt{e}\Lambda^2 = 1$.
The red line shows the arc $|\tau|=1$.}
\label{fig: V2_k8_phi10-3_m10-1_n3.3*10^5}
\end{figure}

As another illustration, we have shown the case $(k_I, \langle \phi \rangle, m_0)=(4, 10^{-3}, 10^{-1})$ and $n/m_{\tau}^4 = 3.3 \times 10^5$ in Fig.~\ref{fig: V2_k8_phi10-3_m10-1_n3.3*10^5}. In this case, we have found $n_c/m_{\tau}^4 \simeq 3.2 \times 10^5$. When $m_{\tau}$ is $10^{17}$GeV, $0.5 \times 10^{17}$GeV, $10^{16}$GeV, the corresponding critical species number is given by $n_{c} \simeq 3.2 \times 10^5, 2.0 \times 10^4, 32$, respectively where we have taken $e^{1/4} \Lambda = 10^{17}$GeV.
As seen in Fig.~\ref{fig: V2_k8_phi10-3_m10-1_n3.3*10^5}, there appears a new vacuum apart from the original vacuum at $\tau=\omega$. We have confirmed that the VEV of $\tau$ at the new vacuum shows $|\delta \tau_*| \sim 0.2$, which is phenomenologically unattractive, and the VEV is almost independent of the strength of the correction characterized by the species number $n$. Thus, we find that the small deviation from the fixed point $\tau=\omega$ is not realized in this parameter region. We will clarify such a behavior analytically in Sec.~\ref{approximation421}. 
In fact, we have numerically found one of the minima at $(x_*, y_*) \simeq (-0.44,0.67)$ in Fig.~\ref{fig: V2_k8_phi10-3_m10-1_n3.3*10^5}. We have found a possibility of CP violation, although within our model, such CP violating vacuum is out of the validity of our approximation. Nevertheless, in general, the radiative correction may cause a CP violating vacuum, of which presence would be highly model dependent, and we will not discuss such a possibility further.

\subsubsection{Approximation}\label{approximation421}
We analytically discuss $V_{\rm eff}$ near the critical point $\tau=\omega$, which clarifies the behavior seen in Figs.\,\ref{fig: V2_k8_phi0.01_m0.1_n_3400_section}-\ref{fig: V2_k8_phi10-3_m10-1_n3.3*10^5}.
Partially expanding Eq.\,(\ref{eq: M^2_1'_expand}) in $\delta\tau$, we obtain
\begin{align}
    \begin{aligned}
\label{eq: approx_V1}
 V_1 
&\simeq \frac{(\xi^2 |\delta \tau|^2+m_0^2)}{32\pi^2} \left[ (\xi^2 |\delta \tau|^2+m_0^2) + 2 \xi^2 |\delta \tau|^2 \delta y \left(\frac{4k_I}{\sqrt{3}}-6\sqrt{3}\right) \right] \log{\left( \frac{\xi^2|\delta \tau|^2 + m_0^2}{\sqrt{e}\Lambda^2} \right)} \\
&\qquad - \frac{(\xi^2 |\delta \tau|^2)^2}{32 \pi^2} \Big[1 + 2\delta y \left(\frac{4k_I}{\sqrt{3}}-6\sqrt{3}\right) \Big] \log{\left( \frac{\xi^2 |\delta \tau|^2}{\sqrt{e}\Lambda^2} \right)},
    \end{aligned}
\end{align}
where we have defined $\xi^2 \equiv \frac{3^{k_I} |D|^2 \langle \phi \rangle^2}{3}$ (see Appendix \ref{appendix: approximation} for its derivation). As expected, $V_1$ is symmetric under $\delta x \rightarrow - \delta x$, which manifests that the local minimum lies on the line $x=-\frac12$. We also find that the value of $V_1$ changes monotonously if we increase or decrease $\delta y$ while keeping $|\delta \tau|$ fixed. 

Using the approximated expression of $V_{\rm eff}$~\eqref{eq: approx_V1}, we discuss the behavior of the possible deviation $\delta\tau$ from the fixed point $\tau=\omega$. In particular, we will take two different parametric regimes $M^2 \gg m_0^2$ and $M^2 \ll m_0^2$. 

For $M^2 \gg m_0^2$ corresponding to ones in Figs.\,\ref{fig: V2_k8_2.2*10^7} and \ref{fig: V2_k8_2.5*10^7}, assuming $\langle \phi \rangle^2 \gg m_0^2$, we are able to further simplify $V_1$ as
\begin{align}
\begin{aligned}
\label{eq: approx_V1_large_M}
V_1\Big|_{x=-1/2}  &\simeq \frac{1}{16\pi^2} m_0^2 \xi^2 \left( \delta y \right)^2 
\left[ 1+ \left( \frac{4k_I}{\sqrt{3}} - 6 \sqrt{3} \right)
\delta y 
\right] 
\log{ \left( \frac{\xi^2 \left( \delta y \right)^2}{\Lambda^2} \right)} \\
&\sim \frac{1}{16\pi^2} m_0^2 \xi^2 \left( \delta y \right)^2 
\log{ \left( \frac{\xi^2 \left( \delta y \right)^2}{\Lambda^2} \right)}.
\end{aligned}
\end{align}
Substituting the extremum condition $\partial_y (V_0 + n V_1)=0$ into the above, the local minimum is found at \begin{equation}
\label{eq: delta_y_1'}
     |\delta y_*|\  \sim \left( 
 \frac{3 \Lambda^2}{3^{k_I} e |D|^2 \langle \phi \rangle^2}
 \right)^{\frac{1}{2}} \exp{\left( - \frac{24 \pi^2 m_{\tau}^4}{3^{k_I} |D|^2 n m_0^2 \langle \phi \rangle^2 } \right)},
\end{equation}
where $\delta y_* = y_* - \frac{\sqrt{3}}{2}$. This analytical expression enables us to estimate the behavior of the deviation $\delta\tau$ as a function of various parameters. In particular, from phenomenological perspectives, $|\delta y|\sim [0.01,0.05]$ results in phenomenologically desired hierarchy of modular forms\cite{Petcov:2022fjf,Kikuchi:2023jap}, and we can estimate the parameters that lead to the desired result with our approximate result~\eqref{eq: delta_y_1'}. To confirm the validity of our approximation, we check the value of $\Delta V_{\rm eff} = |V_{\rm eff}(\omega) - V_{\rm eff}(\omega + i \delta y_*)|$. The linear order expansion yields \begin{equation}
- \Delta  V_{\rm eff} \simeq \delta y_* \cdot \frac{\partial}{\partial y} V_{\rm eff}\Big|_{(\delta x,\delta y)=(0,\frac{\delta y_*}{2})}.
\end{equation} 
With the aid of Eq.\,(\ref{eq: delta_y_1'}), we obtain
\begin{align}
\begin{aligned}
\label{eq: Delta_V_approx}
    \Delta V_{\rm eff} \sim \frac{\log{4}}{16\pi^2 e}\Lambda^2nm_0^2 \ \exp{\left( - \frac{48 \pi^2 m_{\tau}^4}{3^{k_I} |D|^2 nm_0^2\langle \phi \rangle^2 }
    \right)}.
\end{aligned}
\end{align}
In our numerical examples shown in Figs.\,\ref{fig: V2_k8_2.2*10^7} and \ref{fig: V2_k8_2.5*10^7}, we numerically obtain $\Delta V_{\rm eff} \simeq 9.6 \times 10^{-5}$ and $\simeq 2.3 \times 10^{-4}$, respectively, whereas our analytic estimation leads to $\Delta V_{\rm eff} \simeq 1.1 \times 10^{-4}$ and $ \simeq 2.6 \times 10^{-4}$, respectively, which are in good agreement, and confirms the validity of our approximations.

Secondly, let us consider the case $M^2 \ll m_0^2$ corresponding to the one shown in Fig.~\ref{fig: V2_k8_phi10-3_m10-1_n3.3*10^5}. Then the potential $V_{\rm eff}$ is approximated as 
\begin{equation}
V_{\rm eff}  \simeq \frac{n}{32 \pi^2} m_0^4 \log{\left( \frac{m_0^2}{\sqrt{e}\Lambda^2} \right)} + 
m_{\tau}^4 \left( 1 - \frac{n}{n_c} \right) |\delta \tau|^2  - m_{\tau}^4 \frac{n}{n_c} 
\left( \frac{4k_I}{\sqrt{3}} - 6\sqrt{3}
\right) |\delta \tau|^2 \delta y.
\end{equation}
This equation provides us with an approximate behavior at $|\delta \tau| \ll \mathcal{O}(1)$ in Fig.~\ref{fig: V2_k8_phi10-3_m10-1_n3.3*10^5}.

\subsection{$\bf{1}''$}
\begin{figure}[H]
\centering
  \begin{minipage}[b]{0.46\linewidth}
    \centering
 \includegraphics[width=73mm]{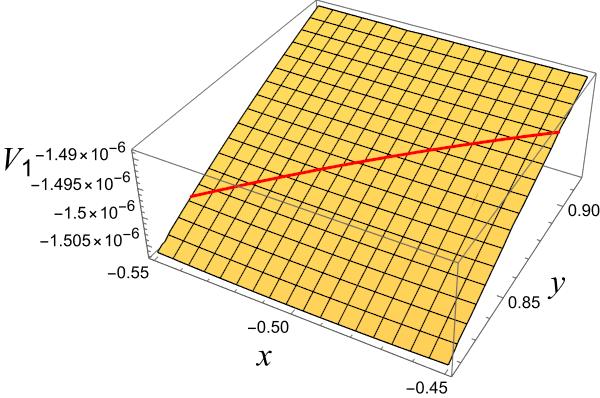}
\caption{The one-loop effective potential $V_1$
in the vicinity of $\tau = \omega$ for $\bm{r}=\bm{1}''$ with the weight $k_Y=8$. The parameters are chosen as $(k_I,\langle \phi \rangle, m_0)=(2, 10^{-3}, 10^{-1})$.
The red line shows the arc $|\tau|=1$. }
\label{V3_k4_phi10-3_m0.1}
  \end{minipage}\quad 
   \begin{minipage}[b]{0.46\linewidth}
    \centering
 \includegraphics[width=73mm]{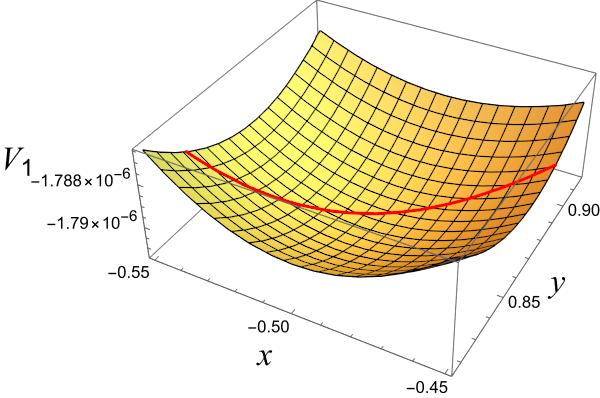}
\caption{The one-loop effective potential $V_1$
in the vicinity of $\tau = \omega$ for $\bm{r}=\bm{1}''$ with the weight $k_Y=8$. The parameters are chosen as $(k_I,\langle \phi \rangle, m_0)=(4, 10^{-3}, 10^{-1})$. The red line shows the arc $|\tau|=1$.}
\label{V3_k8_phi10-3_m0.1}
  \end{minipage}
\end{figure}

Finally, let us consider the case with {$\bm{r}=\bm{1}''$} with weight $k_Y=8$, namely $Y_{\bm{1}''}^{(8)}$. 
In the vicinity of $\tau = \omega$, we have 
\begin{equation}
    M^2 = 3^{k_I} |E|^2 \langle \phi \rangle^2 \left[ 1 +  \frac{4k_I-16}{\sqrt{3}} \delta y - \frac{8}{3} (\delta x)^2 + 
  \frac{2}{3}(4k_I^2 -34k_I +68)(\delta y)^2 + \cdots
    \right],
\end{equation}
from Eq.\,(\ref{eq: Y1''_series})
where $E= Y_{\bm{1}''}(\omega) \simeq -2.05 -3.55i$.
It follows that the first derivative of the 1-loop quantum correction $V_1$ with respect to $x={\rm Re}\tau$ is zero at the fixed point. However, the derivative of $V_1$ with respect to $y={\rm Im}\tau$ does not vanish if $k_I \neq 4$,
\begin{align}
\begin{aligned}
\label{eq: V3_SUSY_deriv}
    \frac{\partial V_1}{\partial x}\bigg|_{\tau = \omega} &= 0, \\
    \frac{\partial V_1}{\partial y}\bigg|_{\tau = \omega} &= \frac{k_I-4}{
    8 \sqrt{3}\pi^2} [M^2\  C(M^2,m_0^2)]_{\tau = \omega}.
\end{aligned}
\end{align}
The second derivatives of $V_1$ at $\tau = \omega$ can be computed as
\begin{align}
\begin{aligned}
\label{eq: V3_SUSY_2nd_deriv}
    \frac{\partial^2 V_1}{\partial x^2}
    \bigg|_{\tau = \omega}   
    &= - \frac{1}{6 \pi^2} [M^2 C(M^2,m_0^2)]_{\tau = \omega}, \\
    \frac{\partial^2 V_1}{\partial y^2}\bigg|_{\tau = \omega}
&= \frac{1}{24\pi^2} \left[ (4k_I^2-34k_I+68) M^2 C(M^2,m_0^2)
      + 8(k_I-4)^2 M^4 \log{\left( 1 + \frac{m_0^2}{M^2} \right)} 
      \right]_{\tau = \omega}, \\
    \frac{\partial^2 V_1}{\partial x \partial y}\bigg|_{\tau = \omega}
    &= 0.
\end{aligned}
\end{align}
If $k_I=4$ which corresponds to the case when $V_1$ is modular invariant\footnote{More precisely speaking, if $k_I\neq 4$ means $\langle \phi \rangle$ has a nontrivial weight under the modular transformation. The VEV $\langle \phi \rangle$ then breaks modular symmetry spontaneously. For $k_I=4$, the modular invariance holds until $\tau$ gets its VEV.}, we find 
$\frac{\partial^2 V_1}{\partial x^2} \big|_{\tau = \omega}   > 0$ and $ \frac{\partial^2 V_1}{\partial y^2} \big|_{\tau = \omega} > 0$,
hence $\tau = \omega$ is a local minimum of $V_1$. One can check this by differentiating $C(M^2,m_0^2)$ with respect to $m_0^2$,
\begin{align}
\frac{\partial C(M^2,m_0^2)}{\partial m_0^2} &= 2 \log{\left( \frac{e (M^2+m_0^2)}{\Lambda^2} \right)},
\end{align}
which is negative 
under the condition $M^2 + m_0^2 \ll \Lambda^2$. Noting that $C(M^2,0)=0$ from Eq.\,(\ref{eq: Curve_C}), we find $C(M^2,m_0^2)< 0$. We show the behavior of $V_1$ of this case in Figs.\,\ref{V3_k4_phi10-3_m0.1} and \ref{V3_k8_phi10-3_m0.1}, where we have chosen $k_I=2$ and $k_I=4$, respectively. We will discuss the case $k_I\neq 4$ in detail in the next sub-subsection.

\subsubsection{Deviation from $\tau = \omega$}
The case $k_I\neq4$ is potentially important for phenomenological applications as $\tau=\omega$ is no longer a minimum of $V_1$, which is a consequence of the spontaneous breaking of the modular symmetry by $\langle \phi\rangle$. Indeed, using Eqs.\,(\ref{eq: V3_SUSY_deriv}) and (\ref{eq: V3_SUSY_2nd_deriv}), the one-loop potential $V_1(\tau)$ is approximated as
\begin{align}
\begin{aligned}
\label{eq: 1''_V1_approx}
V_1(\tau) &\simeq V_1(\omega) + \delta y  
\frac{\partial}{\partial y}V_1(\tau)|_{\tau=\omega}
+ \frac{1}{2} \left( 
(\delta x)^2 \frac{\partial^2}{\partial x^2} + (\delta y)^2 \frac{\partial^2}{\partial y^2}
\right)
V_1 (\tau)|_{\tau=\omega} + \mathcal{O}(|\delta \tau|^3).
    \end{aligned}
\end{align}  
The stationary condition on the total potential 
$V_0+n V_1$ reads
\begin{align}
\begin{aligned}
\label{eq: V3_delta}
\delta x_* &=0, \\
\delta y_* &= - \frac{
\partial_y V_1|_{\tau=\omega}}{2 m_{\tau}^4/n 
{ + }
\partial_y^2 V_1|_{\tau=\omega}}
+ \mathcal{O}(|\delta \tau|^2).
\end{aligned}
\end{align}
Again, $\delta x_*=0$ is a consequence of the CP invariance, and therefore, the minimum in this model does not break CP invariance either. We have shown the deviation $\delta y_*$ as a function of $\langle \phi \rangle$ and $m_0$ in Fig.\,\ref{tb:V3,k=4,deviation} with parameters, $k_I=2$ and $n/m_{\tau}^4=3 \times 10^5$. As is clear from the figure, sufficiently large $\langle \phi\rangle$ and $m_0$ may realize a phenomenologically favored value $|\delta y_*|\sim 0.03$.

\begin{figure}[H]
\centering
 \includegraphics[width=100mm]{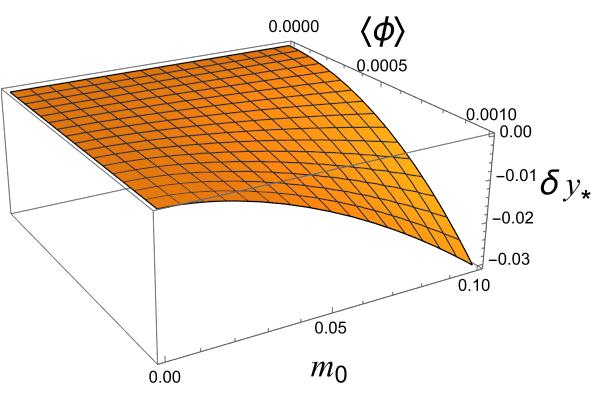}
\caption{The deviation $\delta y_*$ as a function of $\langle \phi \rangle$ and $m_0$. The parameters are chosen as $k_I=2$ and $ n/m_{\tau}^4 = 3\times 10^5$.}
\label{tb:V3,k=4,deviation}
\end{figure}

\section{Conclusion}
\label{conclusion}
We have studied the 1-loop effective potential within $A_4$ modular flavor symmetric SUSY models and its application to the stabilization of the complex structure modulus $\tau$. In particular, we have focused on the models in which $A_4$ modular forms have $k_Y=8$ and belong to one of the singlet representations $\bm r=\bm 1,\bm 1',\bm 1''$. The 1-loop effective potential originates from supermultiplets $\Phi_I$. For generality of our analysis, we have not specified the origin of the soft SUSY breaking and parametrized its strength by the soft scalar mass parameter $m_0$. As expected the resultant one-loop potential is always proportional to $m_0^2$ as it vanishes in the SUSY limit $m_0\to 0$. 

In our analysis, we have been concerned with an $ST$ invariant fixed point $\tau=\omega$ at which residual $Z_3$ symmetry exists. Such a fixed point is phenomenologically interesting as the slightly broken $Z_3$ symmetry naturally realizes the hierarchical flavor structures of the standard model. Thus, we have focused particularly on whether the 1-loop correction can lead to the small deviation from the fixed point by assuming the ``tree level'' potential $V_0$ of a simple form~\eqref{V0form}.

We have found that depending on the choice of the representation $\bm r=\bm1,\bm 1', \bm 1''$ of the $A_4$ modular forms, the resulting 1-loop effective potential shows different behaviors and accordingly the different minima for each case. For a trivial singlet choice $\bm r=\bm 1$, $V_1$ turns out to be flat near the fixed point $\tau=\omega$, and the desired deviation cannot be realized. Nevertheless, we have found the possibilities to realize the phenomenologically important (small) deviation $\delta\tau$ for $\bm r=\bm 1', \bm 1 ''$. For the former case, we have found that a large number of flavors contributing to the effective potential are necessary to make the one-loop contribution compatible with the ``tree level'' potential when the modulus mass $m_\tau$ is as large as the compactification scale.\footnote{Generic string compactification leads to a large number of massless modes at the compactification scale \cite{Ibanez:2012zz}.} We here emphasize that this requirement does not conflict with the perturbative treatment of the theory as the ``tree level'' potential here has nothing to do with the ``Yukawa couplings'' of $\Phi_I$. Nevertheless, one must be careful about the introduction of too large numbers of species that leads to the breakdown of the perturbative description of (quantum) gravity. 
When the modulus mass is lighter than the compactification scale, 
a small number of flavors is enough. For $\bm r= \bm 1 ''$ case, it turns out that the (spontaneous) modular symmetry breaking situation $k_I\neq4$ makes $V_1$ to be non-stationary at $\tau=\omega$, which leads to a small deviation of the minimum from $\tau=\omega$, whereas the modular symmetric choice $k_I=4$ fails to realize a small deviation of the minimum.

We would like to stress that, generally speaking, the modulus originates from a gravitational degrees of freedom, which very weakly couples to matter sector. Therefore, the (small) deviation from the tree level vacuum requires both sufficiently large $\langle \phi\rangle$ and the number of species $n$ that strengthen the loop correction as we have seen more explicitly. One of the lessons from our work is that only when the ``tree level'' potential is sufficiently small, the one-loop effective potential may lead to a phenomenologically desired deviation otherwise extremely large amount of species or too large VEV $\langle \phi \rangle$  is required, which would be in conflict with the effective field theory descriptions. Nevertheless, such a situation is ubiquitous in string theory since moduli fields are generically light unless $p$-form fluxes are introduced. Therefore, relatively small tree level/non-perturbative potential can naturally be realized, and then the loop contribution can compete with it, which may results in the phenomenologically realistic vacuum as we have shown in this work.

The radiative corrections would also have some impacts on the dynamical moduli trapping mechanism~\cite{Kofman:2004yc}, by which moduli fields can be trapped at the points that matter fields become massless and symmetries are enhanced. In our previous study~\cite{Kikuchi:2023uqo}, we have shown that the complex structure modulus can be trapped at $\tau=\omega$ under the assumption that the 1-loop effective potential from matters is canceled by SUSY. However, in realistic models, SUSY should be spontaneously broken and the 1-loop effective potential arises as in this work. It would be important to study the effect of the 1-loop effective potential to the moduli trapping mechanism, which enables us to answer whether the complex structure modulus can be stabilized even if the modulus is not stabilized in the very early universe. We will leave such issues in future work.

\vspace{1.5 cm}
\noindent
{\large\bf Acknowledgement}\\
This work was supported by 
 JSPS KAKENHI Grant Numbers 
JP23K03375 (TK), JST SPRING Grant Number JPMJSP2119 (KN), and Waseda University Grant for Special Research Projects (Project number: 2023C-584) (YY).


\appendix
\section{Group theoretical aspects of $A_4$}
\label{appendix: group_A4}
The $A_4$ group has two generators $S$ and $T$ satisfying the following algebraic relations:
\begin{align}
  S^2 = (ST)^3 = T^3 = {\mathbb I}.
\end{align}
Four irreducible representations exist in $A_4$, which are three singlets $\bf{1}$, $\bf{1}'$ and $\bf{1}''$ and one triplet $\bf{3}$.
Each irreducible representation is given by
\begin{align}
\begin{aligned}
  &{\bm{1}} \quad \rho(S)=1, ~\rho(T)=1, \\
  &\bm{1}' \quad \rho(S)=1,~\rho(T)=\omega, \\
  &\bm{1}'' \quad \rho(S)=1, ~\rho(T)=\omega^2, \\
  &\bm{3} \quad 
  \rho(S) = \frac{1}{3}
  \begin{pmatrix}
    -1 & 2 & 2 \\
    2 & -1 & 2 \\
    2 & 2 & -1 \\
  \end{pmatrix},\quad
  \rho(T) =
  \begin{pmatrix}
    1 & 0 & 0 \\
    0 & \omega & 0 \\
    0 & 0 & \omega^2 \\
  \end{pmatrix},
\end{aligned}
\end{align}
where we adopted the $T$-diagonal basis.
Their multiplication rules are shown in Table \ref{tab:MultiRuleinA4}.
\begin{table}[H]
\begin{center}
\renewcommand{\arraystretch}{1}
\begin{tabular}{c|c} \hline
  Tensor product  & $T$-diagonal basis \\ \hline
  $\bm{1}'' \otimes \bm{1}'' = \bm{1}'$ & \multirow{3}{*}{$a^1b^1$} \\
  $~~~~~~~~~~\bm{1}' \otimes \bm{1}' = \bm{1}''$ ~~$(a^1 b^1)$ & \\
  $\bm{1}'' \otimes \bm{1}' = \bm{1}$ & \\ \hline
  \multirow{2}{*}{$\bm{1}'' \otimes \bm{3} = \bm{3}$ ~~$(a^1 b^i)$} & \multirow{2}{*}{$\left(\begin{smallmatrix} a^1b^3\\ a^1b^1\\ a^1b^2\\\end{smallmatrix}\right)$} \\
  & \\ \hline
  \multirow{2}{*}{$\bm{1}' \otimes \bm{3} = \bm{3}$~~$(a^1 b^i)$} & \multirow{2}{*}{$\left(\begin{smallmatrix} a^1b^2\\ a^1b^3\\ a^1b^1\\ \end{smallmatrix}\right)$} \\
  & \\ \hline
  \multirow{5}{*}{$\bm{3}\otimes \bm{3}=\bm{1}\oplus 
  {
  \bm{1}' \oplus \bm{1}'' }
  \oplus \bm{3} \oplus \bm{3}$} & $\begin{smallmatrix}(a^1b^1+a^2b^3+a^3b^2)\end{smallmatrix}$ \\
  & $\oplus\begin{smallmatrix}(a^1b^2+a^2b^1+a^3b^3)\end{smallmatrix}$ \\
  & $\oplus\begin{smallmatrix}(a^1b^3+a^2b^2+a^3b^1)\end{smallmatrix}$ \\
 \multirow{2}{*}{$(a^ib^j)$}  & \multirow{2}{*}{$\oplus\frac{1}{3}\left(\begin{smallmatrix} 2a^1b^1-a^2b^3-a^3b^2\\ -a^1b^2-a^2b^1+2a^3b^3\\ -a^1b^3+2a^2b^2-a^3b^1\\ \end{smallmatrix}\right)$} \\
  & \\
  & \multirow{2}{*}{$\oplus\frac{1}{2}\left(\begin{smallmatrix} a^2b^3-a^3b^2\\ a^1b^2-a^2b^1\\ -a^1b^3+a^3b^1\\ \end{smallmatrix}\right)$} \\
  & \\ \hline
\end{tabular}
\end{center}
\caption{Multiplication rule in irreducible representations of $A_4$.}
\label{tab:MultiRuleinA4}
\end{table}
Further details are explained in \cite{Ishimori:2010au,Kobayashi:2022moq}.

\section{$A_4$ modular forms}
\label{sec: A_4}
The modular forms of $A_4$ with even weights can be written in terms of the Dedekind eta function $\eta(\tau)$ and its derivative,
\begin{align}
  &\eta(\tau) = q^{1/24} \prod_{n=1}^{\infty} (1-q^n), \quad q = e^{2\pi i\tau}, \\
  &\eta'(\tau) \equiv \frac{d}{d\tau} \eta(\tau).
\end{align}
Modular forms of weight 2 which transform as a triplet $\bf{3}$ of $A_4$ can be given by \cite{Feruglio:2017spp}
\begin{equation}
Y_{\bf 3}^{(2)}(\tau) = 
\begin{pmatrix}
    Y_1 \\ Y_2 \\ Y_3
\end{pmatrix},
\end{equation}
where 
\begin{align}
\begin{aligned}
 Y_1(\tau) &= \frac{i}{2\pi} \left( \frac{\eta'(\tau/3)}{\eta(\tau/3)} + \frac{\eta'((\tau+1)/3)}{\eta((\tau+1)/3)} +  \frac{\eta'((\tau+2)/3)}{\eta((\tau+2)/3)}
-\frac{27\eta'(3\tau)}{\eta(3 \tau)}\right), \\
Y_2(\tau) &= \frac{-i}{\pi} \left( \frac{\eta'(\tau/3)}{\eta(\tau/3)} + \omega^2 \frac{\eta'((\tau+1)/3)}{\eta((\tau+1)/3)} + \omega \frac{\eta'((\tau+2)/3)}{\eta((\tau+2)/3)}
\right),  \\
Y_3(\tau) &= \frac{-i}{\pi} \left( \frac{\eta'(\tau/3)}{\eta(\tau/3)} + \omega \frac{\eta'((\tau+1)/3)}{\eta((\tau+1)/3)} + \omega^2 \frac{\eta'((\tau+2)/3)}{\eta((\tau+2)/3)}
\right).
\end{aligned}
\end{align}
Modular forms with higher weights can be constructed by taking products of $Y_{\bf{3}}^{(2)}$. For example, singlet modular forms of weight $8$ can be constructed as 
\begin{align}
\begin{aligned}
Y_{\bf 1}^{(8)}&=(Y_1^2 + 2Y_2 Y_3)^2, \\
Y_{{\bf 1}'}^{(8)}&=(Y_1^2 + 2Y_2 Y_3)(Y_3^2 + 2Y_1 Y_2), \\
Y_{{\bf 1}''}^{(8)}&=(Y_3^2 + 2Y_1 Y_2)^2.
\end{aligned}
\end{align}

\section{Approximation}
\label{appendix: approximation}
From Eq.\,(\ref{eq: M^2_1'_expand}), we approximate
\begin{equation}
    (M^2 + m_0^2)^2 = (\xi^2 |\delta \tau|^2 + 
    m_0^2) 
    {
    \left[ 
    (\xi^2 |\delta \tau|^2 + m_0^2) +2 \xi^2 |\delta \tau|^2 \delta y \left(\frac{4k_I}{\sqrt{3}}-6\sqrt{3}\right)
    + \mathcal{O}(|\delta \tau|^4)
     \right]
     },
\end{equation}
and 
\begin{align}
\begin{aligned}
    \log{\left( \frac{M^2+m_0^2}{\sqrt{e} \Lambda^2} \right)} &\simeq \log{\left[ 
    \frac{\xi^2 |\delta \tau|^2 \left( 1 + \delta y \left(\frac{4k_I}{\sqrt{3}}-6\sqrt{3} \right) 
    \right) + m_0^2}{\sqrt{e} \Lambda^2} 
    \right]} \\
    &= \log{\left( 
    \frac{\xi^2 |\delta \tau|^2  + m_0^2}{\sqrt{e} \Lambda^2} 
    \right)} + \log{\left(1+\frac{\xi^2 |\delta \tau|^2 \delta y {\left(\frac{4k_I}{\sqrt{3}}-6\sqrt{3} \right) }}{\xi^2 |\delta \tau|^2 + m_0^2}
    \right)}.
\end{aligned}
\end{align}
The second term is of order $|\delta \tau|$ and is negligible compared with the first term.
Thus, we obtain 
\begin{align}
\begin{aligned}
    &(M^2+m_0^2)^2 \log{\left( \frac{M^2+m_0^2}{\sqrt{e} \Lambda^2} \right)}\\
    & \simeq (\xi^2 |\delta \tau|^2+m_0^2) \left[ (\xi^2 |\delta \tau|^2+m_0^2) + 2 \xi^2 |\delta \tau|^2 \delta y \left(\frac{4k_I}{\sqrt{3}}-6\sqrt{3}\right) \right] \log{\left( \frac{\xi^2|\delta \tau|^2 + m_0^2}{\sqrt{e} \Lambda^2} \right)}.
    \end{aligned}
\end{align}
This leads to Eq.\,(\ref{eq: approx_V1}).


 \end{document}